\documentclass[12pt]{iopart}
\usepackage{iopams}
\usepackage{graphics}
\usepackage[next]{inputenc}
\usepackage[dvips]{epsfig}
\def\be{\begin{equation}}
\def\ee{\end{equation}}

\def\bi{\begin{itemize}}
\def\ei{\end{itemize}}
\def\bn{\begin{enumerate}}
\def\en{\end{enumerate}}
\def\bea{\begin{eqnarray}}
\def\eea{\end{eqnarray}}
\def\no{\nonumber}
\def\ba{\begin{array}}
\def\ea{\end{array}}
\def\bd{\begin{displaymath}}
\def\ed{\end{displaymath}}
\def\la{\langle}
\def\ra{\rangle}

\begin{document}

\title{Entanglement in heterogeneous spin-$(1, \frac 12)$ and homogeneous spin-$1$ systems}

\author{N. Askari$^1$ and J. Abouie $^{2,3}$}

\address{$^{1}$Department of Physics, Shahrood University of Technology, Shahrood 36199-95161,
Iran}
\address{$^{2}$ Department of Physics, Institute for Advanced Studies in Basic Sciences, Zanjan 45137-66731, Iran}
\address{$^{3}$ School of physics, Institute for Research in Fundamental Sciences (IPM),
Tehran 19395-5531, Iran}

\ead{jahan@iasbs.ac.ir}
\date{\today}

\begin{abstract}
We study the bipartite entanglement of two general classes of heterogeneous
spin-($1,\frac 12$) and homogeneous spin-$1$ systems. By employing the spin correlation functions, we obtain the reduced two-spin density matrix (DM) and the negativity for these two classes of quantum spin models. We show explicitly that in addition to the one and two-point correlations, the triad and quad correlations ($t_{\alpha\beta}^{\delta}=\la S_{\alpha}S_{\beta}s_{\delta}\ra$ and $q_{\alpha\beta}^{\delta\gamma}=\la S_\alpha S_\beta s_\delta s_\gamma\ra$ where $\alpha, \beta, \delta, \gamma=\pm, z$) play crucial role in the bipartite entanglement between spins $s>1/2$. These correlations represent the spin $\frac 12$-quadrupole and quadrupole-quadrupole correlations, respectively. 
These correlations do not appear in the spin-$\frac 12$ models.
Our results are general and applicable to the different several models of interest with higher reflectional, translational,
spin-flip and $U(1)$ symmetries. The entanglement of many attractive models
are investigated.
\end{abstract}


\maketitle
\section{Introduction}
In many-body physics, quantum spin systems have received a lot of attentions due to their importance in the description of different phenomena such as antiferromagnetic Mott insulating behavior,
Bose-Einstein condensation (BEC) \cite{BEC}, spin supersolidity \cite{Seng07} and gapless L\"{u}ttinger liquid physics.\cite{Giam04}
Apart from pure scientific interest, spin systems possess a potential for various technological applications
such as nano-technology and quantum computations.
They are prototype realization of many relevant properties of quantum implementation in advance technologies such as fabrication of spin transverse torque switching devices.\cite{device}
Therefore different aspects of a quantum phase are of utmost importance for scientists and engineers.

Entanglement is a kind of quantum nonlocal correlation which is used as an indicator of quantum phase transitions.\cite{Oste02, Vida03} The collective behavior of a system is associated with the development of this nonlocal correlations as well as classical correlations. Moreover condensed matter systems can be efficiently simulated using techniques related to entanglement.\cite{Vidal} The usefulness of entanglement in condensed matter physics leads us to study it in the parity invariant (PI) quantum spin systems. Recently, the geometrical entanglement has been studied for one-dimensional systems with space or space-time inversion symmetries.\cite{Orus11}
The main reasons to consider the systems with parity symmetry are as follows: (i) The external parameters such as single site and lattice anisotropies, spin-orbit couplings and external magnetic and electric
fields do not break the parity and we may investigate the simultaneous effects of these parameters in the entanglement properties of the state. (ii) It is a general model and the results are employed to study the entanglement properties of many interesting systems with higher reflectional, translational, spin-flip and U(1) symmetries.

In the present paper, we investigate the bipartite entanglement in parity invariant(PI) heterogeneous mixed spin-($1,\frac 12$) and homogeneous spin-$1$ systems. Many theoretical and experimental investigations of the physical properties of $S>\frac 12$ spin systems have been done since Haldane's conjecture\cite{Haldane} and synthesis of bimetallic molecular magnets and supramolecular composite systems.\cite{Kahn93} An interesting system with parity symmetry is the XXZ spin-($a,\frac 12$) or spin-$1$ model in a transverse magnetic field with Dzyaloshinskii-Moriya interaction ($\overrightarrow{D}\cdot\overrightarrow{S}_i\times\overrightarrow{S}_j$, arising from spin-orbit coupling).
We obtain an expression for the two-spin reduced density matrix $\rho^{(2)}$ (DM) and the negativity in terms of the spin correlation functions. All the information needed to analyze bipartite entanglement is contained in the two-spin reduced
density matrix, obtained from the wave function of the state after all the spins except those at positions $i, j$
have been traced out. The resulting reduced density matrix represents a mixed state of a bipartite system for which a good deal of work has been devoted to quantify its entanglement.
The spin correlation functions show the collective critical behaviors of particles, as well as their ordering and fluctuations in a system.
Utilizing spin-spin correlation functions, enables us to investigate the effects of various external parameters (magnetic and electric field, anisotropies and temperature) to study the entanglement properties of a state such as the dynamics of entanglement, entanglement witnesses and quantum fidelity.
We show explicitly that in addition to the magnetizations $\la S_i^{\alpha}\ra$ and two-point correlations $\la S_i^{\alpha}S_j^{\alpha}\ra$, the information of triad and quad correlations ($t_{\alpha\beta}^{\delta}=\la S_{\alpha}S_{\beta}s_{\delta}\ra$ and $q_{\alpha\beta}^{\delta\gamma}=\la S_\alpha S_\beta s_\delta s_\gamma\ra$ where $\alpha, \beta, \delta, \gamma=\pm, z$) play key role in the bipartite entanglement in ferrimagnetic $(S=1, s=\frac 12)$ and antiferromagnetic $S=1$ spin systems. They are indeed the spin $\frac 12$-quadrupole and quadrupole-quadrupole correlations.
Finally, employing our results we study the entanglement in different several models of interest with higher symmetries.

\section{PI heterogeneous ($S=1, s=\frac 12$) spin systems}

A parity invariant state satisfies the relation $[{\cal P}, \rho]=0$
where the parity operator ${\cal P}$ is defined as: 
\begin{equation}
{\cal P}=e^{-i\pi\sum_{i,j}(S_i^z+s_j^z)}=\prod_{i,j}(1-(S_i^z)^2)s_j^z.
\end{equation}
For the homogeneus spin-$\frac 12$ system the parity operator is reduced to the more familiar form: $\prod_i s_i^z$.
Let us utilize spin correlation functions to obtain the elements of DM.
A relationship between concurrence\cite{Woot98} and two-point correlation functions of the spin-$\frac 12$ anisotropic Heisenberg $XY$ model in the presence of a transverse magnetic field has been already presented.\cite{Amico04}
For any spin-$\frac 12$ system, all elements of DM are acquired by using the magnetizations and the two-point correlations. However, these functions are not sufficient to attain the DM of PI spin-($1, \frac 12$) or spin-1 systems. As a matter of fact the quadrupolar effects of the spin-1 particles play important role on the physical behavior of the system and we should employ them to achieve the DM. Defining the triad correlations as $t_{\alpha\beta}^{\delta}=\la S_{\alpha}S_{\beta}s_{\delta}\ra$ where $\alpha, \beta, \delta=\pm, z$
and working in the basis of factorized state
the DM of a PI state is given by

\begin{equation}
\rho^{(2)}_{S=1,s=\frac 12}=\left(\begin{array}{cccccc}
\frac 14 t_{zz^+}^{z^+} & 0 & \frac 14 t_{--}^{z^+}& 0 & \frac{1}{\sqrt 2}t_{z-}^{-}& 0\\
 0&\frac 12 t_{z^+z^-}^{z^+}&0&\frac{1}{\sqrt 2}t_{z+}^{-}&0&-\frac{1}{\sqrt 2}t_{z-}^{-}\\
\frac 14 t_{++}^{z^+}&0&-\frac 14 t_{zz^-}^{z^+}&0&-\frac{1}{\sqrt 2}t_{z+}^{-}&0\\
0&\frac{1}{\sqrt 2}t_{-z}^{+}&0&\frac 14 t_{zz^+}^{z^-}&0&\frac 14 t_{--}^{z^-}\\
\frac{1}{\sqrt 2}t_{+z}^{+}&0&-\frac{1}{\sqrt 2}t_{-z}^{+}&0&\frac 12 t_{z^+z^-}^{z^-}&0\\
0&-\frac{1}{\sqrt 2}t_{+z}^{+}&0&\frac 14 t_{++}^{z^-}&0&-\frac 14 t_{zz^-}^{z^-}\\
\end{array}\right)
\label{DM1-1/2}
\end{equation}
where
\begin{eqnarray}
&\no t_{\alpha\beta}^{z^{\pm}}=\la S_{\alpha}S_{\beta}[1\pm 2s_z]\ra=\la S_{\alpha}S_{\beta}\ra\pm 2t_{\alpha\beta}^z,\\
&\no t_{z^{\pm}\beta}^{\delta}=\la [1\pm S_z]S_{\beta}s_{\delta}\ra=g_{\beta\delta}\pm t_{z\beta}^{\delta}, ~~~~g_{\beta\delta}=\la S_{\beta}s_{\delta}\ra.
\label{triad}
\end{eqnarray}
Some off-diagonal elements in (\ref{DM1-1/2}) like $t_{z\pm}^{-}=\langle S_zS_{\pm}s_-\rangle$ and $t_{\pm z}^{+}=\langle S_{\pm}S_zs_+\rangle$ are the correlations of the transverse component of spin-$\frac 12$ with operators $(S_zS_{\pm})$ and $(S_{\pm}S_z)$.
These are nothing but the quadrupolar operators\cite{quad} which have been introduced to describe the quadrupolar order in the spin-$1$ bilinear-biquadratic Heisenberg model.\cite{Lauc06} The reduced density matrix strongly depends on the correlations between the spin-$\frac 12$ particle and the quadrupoles. Some of these correlations are suppressed at the presence of other symmetries.
For example, for states with parity and spin-flip symmetries\cite{spin-flip}
the equalities $\Re t_{+z}^{+}=\Re t_{-z}^{+}=0$ ($\Re$ denotes the real part), $t_{++}^{z^-}=t_{--}^{z^+}$ and $M_z=m_z=t_{zz}^{z}=0$ are gained. ($M_z=\la S_z\ra$ and $m_z=\la s_z\ra$
are the sub-lattice magnetizations in $z$ direction). Moreover, adding $U(1)$ symmetry to the parity,
the equalities $t_{--}^{z^+}=t_{--}^{z^-}=t_{z-}^{-}=0$ are obtained.
By making use of vigorous analytical and numerical approaches such as bosonization, Green's functions techniquw, exact diagonalization Lanczos
and density matrix renormalization group methods we can study the behaviors of the correlation functions.

{\it Negativity for spin-(1,1/2) system}

\noindent
Utilizing the reduced density matrix (\ref{DM1-1/2}) we can obtain the entanglement, as quantified by negativity and relative entropy of entanglement, between two spins in a PI spin-$(1,\frac 12)$ state. In this paper we focus on the negativity and the relative entropy of entanglement is left for the future.
Negativity is based on the trace norm of the partial transpose of density matrix and is given by the formula:
\begin{equation}
\mathcal{N}_{12}(\rho)\equiv\frac{\Vert\rho^{T_{1}}\Vert_1-1}{2},
\end{equation}
which corresponds to the absolute value of the sum of the negative eigenvalues of partial transpose density matrix, $\rho^{T_{1}}$.\cite{Vidal02}

\begin{figure}
\centerline{
\includegraphics[width=9cm,angle=0]{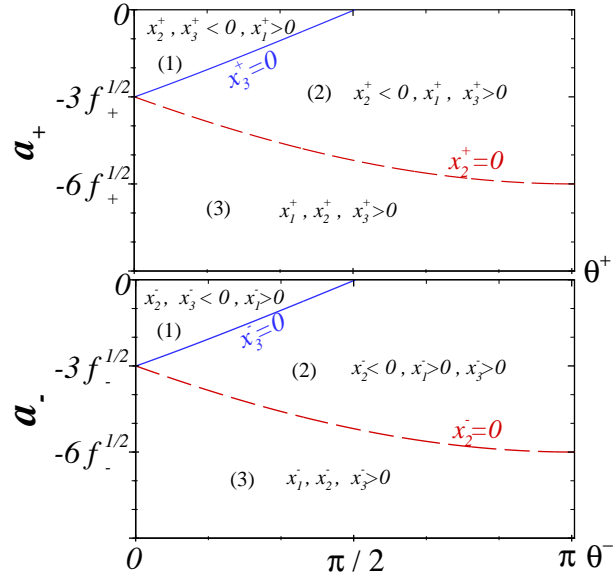}}
\caption{Different possibilities for the sign of the solutions of the two cubic equations.
By Comparing $-3\sqrt{f_{\pm}}$ and $-6\sqrt{f_{\pm}}$ with $1$ we find the allowable regions simply.
} \label{Neg-val}
\end{figure}

The negative eigenvalues are obtained by solving two cubic equations: $x^{3} +a_\pm x^{2}+b_\pm x+c_\pm=0$
where $a_\pm=-\frac 12\pm(m_z-2t_{zz}^{z})$ are bounded by $-1\leq a_{\pm}\leq 0$ and
\begin{eqnarray}
&\no c_\pm=\frac{1}{32}t_{z^+z^-}^{z^{\mp}}\big[(M_z\pm 2 g_{zz})^2
-[t_{zz}^{z^\pm}]^2+|t_{--}^{z^{\pm}}|^2\big]\\
&\no+\frac 18\big[t_{z-}^{-}(t_{++}^{z^\pm}t_{z-}^{+}\pm t_{zz^\pm}^{z^\pm}t_{z+}^{+})+t_{z+}^{-}(t_{--}^{z^\pm}t_{z+}^{+}\mp t_{zz^\mp}^{z^\pm}t_{z-}^{+})\big],\\
&\no b_\pm=\frac{1}{16}\big\{t_{zz}^{z^\pm}(t_{zz}^{z^\pm}+4t_{z^+z^-}^{z^\mp})
-(M_z\pm 2g_{zz})^2
-|t_{--}^{z^\pm}|^2-8[|t_{z+}^-|^2+|t_{z-}^-|^2]\big\}\label{cubic}.
 \end{eqnarray}
By defining the parameters $f_\pm=\frac 19a_\pm^2-\frac 13b_\pm$, $r_\pm=\frac 16(b_\pm a_\pm-3c_\pm)-\frac{1}{27}a_\pm^3,
$ and $\cos\theta_\pm=r_\pm f_\pm^{-\frac 32}$ the standard solutions of the cubic equations are
\begin{eqnarray}
\no &x_1^{\pm}=2\sqrt{f_\pm}\cos(\frac{\theta_\pm}{3})-\frac{a_\pm}{3},\\ \no&x_2^{\pm}=2\sqrt{f_\pm}\cos(\frac{\theta_\pm+2\pi}{3})-\frac{a_\pm}{3},\\ &x_3^{\pm}=2\sqrt{f_\pm}\cos(\frac{\theta_\pm+4\pi}{3})-\frac{a_\pm}{3},
\end{eqnarray}
where $\theta_{\pm}\in[0,\pi]$.
The negativity is the absolute value of the sum of the negative $ x$'s.
All possibilities for the sign of eigenvalues have been explicitly shown in Fig. (\ref{Neg-val}).
If $(x_i^+, x_i^-)\in$ region (1), then ${\cal N}_{12}=\sum_{i=2,3}|x_i^+|^2+|x_i^-|^2$ and the state is maximally entangled. However in region (3) negativity is exactly zero which corresponds to a fully factorized state and the state is partially entangled in other cases.
A system with parity symmetry is the ferrimagnetic XXZ spin-$(1,\frac 12)$ model in a transverse magnetic field with Dzyaloshinskii-Moriya interaction ($\overrightarrow{D}\cdot\overrightarrow{S}_i\times\overrightarrow{s}_j$). The parameters $\theta_\pm$ and $f_\pm$ depend on the transverse magnetic field and anisotropy parameter $D$. Quantum phases are specified by $D$ and magnetic field and the negativity would give useful information on the entanglement of different phases of the system.

To see more advantages of these results it is worth to investigate the entanglement of states with higher symmetries. By adding the spin-flip symmetry to the parity, the coefficients of cubic equations are simplified as $a=a_+=a_-, b=b_+=b_-, c=c_+=c_-$ and
the negative eigenvalues have been indicated in Fig.(\ref{Neg-val}).
For $b<0$, the state is always entangled and the entanglement depends on $\theta$. For small values of $\theta$ the state could be maximally entangled and the entanglement is decreased by increasing $\theta$.
At $b=0$ the entanglement is independent of $\theta$ and the state is always partially entangled. For $0<b<\frac{1}{16}$ and small values of $\theta$ the state is fully factorized however it is partially entangled by increasing $\theta$. Finally for $b>\frac{1}{16}$ the state is always fully factorized.
Some examples of models with such symmetries are as follows: (i)
Mixed spin XXZ and XY systems in a transverse magnetic field\cite{Langari11},
and (ii) Mixed spin Ising model with (or without) next nearest neighbor interaction in the presence of a transverse and a longitudinal field.
Ferrimagnetic Ising models have been studied with a renewed interest, especially in connection with compensation points. These points occur at temperatures below the critical temperature, at which the sublattice magnetizations cancel exactly, and the total magnetization changes sign.
These systems maybe used in technical applications, most notably in magnetic recording.

Finally a mixed state with spin-flip, $U(1)$ and parity symmetries is entangled if
$\frac 14[1-\frac 12\la S_z^2\ra- g_{zz}-\sqrt{[\frac 32\la S_z^2\ra+g_{zz}-1]^2-8[t_{z+}^-]^2}]<0$.
All anisotropic bond-alternating spin-($1,\frac 12$) ferrimagnets (with dimerization, trimerization and tetramerization) in a longitudinal field, the ferrimagnetic XXZ models in a longitudinal field have these types of symmetries.
The results are applicable to the synthesized ferrimagnetic compounds with Ising-like anisotropy such as $CoNi(EDTA)6H_2O$ where $(S,s)=(1(Ni),\frac 12(Co))$.\cite{Drill85}
\section{PI homogeneous $S=1$ spin systems}

In this section we obtain an expression for the reduced density matrix of PI spin-$1$ systems. The parity operator is simplified as ${\cal P}=\prod_i(1-2(S_i^z)^2)$. In addition to the sublattice magnetizations and two-point correlations other correlations should be employed to find all elements of the DM. By introducing the quad correlations $q_{\alpha\beta}^{\delta\gamma}=\la S_\alpha S_\beta s_\delta s_\gamma\ra$ (indices are $\pm$ and $z$) and working in the basis of factorized state
the DM is given by
\begin{eqnarray}
\no\rho^{(2)}_{_{S=s=1}}=\frac 12\times\\
\no\left(\begin{array}{ccccccccc}
\frac{1}{2}q_{zz^+}^{zz^+}&0&\frac{1}{2}q_{zz^+}^{--}&0&q_{z-}^{z-}&0&\frac 12 q_{--}^{zz^+}&0&\frac 12 q_{--}^{--}\\
0&q_{zz^+}^{z^+z^-}&0&q_{z-}^{+z}&0&-q_{z-}^{-z}&0&q_{--}^{z^+z^-}&0\\
\frac{1}{2}q_{zz^+}^{++}&0&-\frac{1}{2}q_{zz^+}^{zz^-}&0&-q_{z-}^{z+}&0&
\frac{1}{2}q_{--}^{++}
&0&-\frac{1}{2}q_{--}^{zz^-}\\
0&q_{z+}^{-z}&0&q_{z^+z^-}^{zz^+}&0&q_{z^+z^-}^{--}&0&-q_{z-}^{-z}&0\\
q_{z+}^{z+}&0&-q_{z+}^{z-}&0&2 q_{z^+z^-}^{z^+z^-}&0&-q_{z-}^{z+}&0&q_{z-}^{z-}\\
0&-q_{z+}^{+z}&0&q_{z^+z^-}^{++}&0&-q_{z^+z^-}^{zz^-}&0&q_{z-}^{+z}&0\\
\frac 12 q_{++}^{zz^+}&0&\frac{1}{2}q_{++}^{--}&0&-q_{z+}^{z-}&0&-\frac{1}{2}
q_{zz^-}^{zz^+}&0&-\frac{1}{2}q_{z^-z}^{--}\\
0&q_{++}^{z^+z^-}&0&-q_{z+}^{+z}&0&q_{z+}^{-z}&0&-q_{zz^-}^{z^+z^-}&0\\
\frac 12 q_{++}^{++}&0&-\frac{1}{2}q_{++}^{zz^-}&0&q_{z+}^{z+}&0&-\frac{1}{2}
q_{z^-z}^{++}&0&\frac{1}{2}q_{zz^-}^{zz^-}\\
\end{array}\right)\\
\label{DMs1}
\end{eqnarray}
where $|S|=|s|=1$ and
\begin{eqnarray}
\no&q_{\alpha\beta}^{\delta z^\pm}=\la S_\alpha S_\beta s_\delta(1\pm s_z) \ra=t_{\alpha\beta}^{\delta}\pm q_{\alpha\beta}^{\delta z},\\
&q_{\alpha z^\pm}^{\delta\gamma}=\la S_\alpha(1\pm S_z)s_\delta s_\gamma\ra=t_{\alpha}^{\delta \gamma}\pm q_{\alpha z}^{\delta\gamma}.
\label{quartic}
\end{eqnarray}
Appearance of the quad correlations like $q_{z-}^{z\pm}=\langle S_zS_-s_zs_{\pm}\rangle$ and $q_{z-}^{\pm z}=\langle S_zS_-s_{\pm}s_z\rangle$ in Eq.(\ref{DMs1}) is noticeable. By writing $S_zS_-$ and $s_zs_{\pm}$ or $s_{\pm}s_z$ in terms of quadrupolar operators\cite{quad} we find that the quadrupole-quadrupole correlations play an essential role in the DM of homogeneous spin-$1$ models.
This kind of correlations appear neither in heterogenous spin-$(1,\frac 12)$ systems nor in homogeneous spin-$\frac 12$ models. They are the characteristic behavior of the spin-$1$ systems.
Adding further symmetries to the parity, suppresses some of such correlations.

Comparing the DM of three spin-$\frac 12$, spin-$(1,\frac 12)$ and spin-$1$ systems, we conclude that always one component of spin-$\frac 12$ operator, and two components of spin-$1$ operator contribute to the triad and quad correlation functions. Moreover, all information for the bipartite entanglement of a spin-$\frac 12$ system are gathered from the magnetizations, staggered magnetizations and the two-point correlations, however in a heterogenous spin-$(1,\frac 12)$ and a homogeneous spin-$1$ system in addition to one and two-point correlations the triad and quad correlations play key roles in the bipartite entanglement of PI states.

{\it Negativity for spin-1 system}- For systems with dimension higher than $2\otimes 3$, inseparable states with positive partial transpose (PPT) exist, i.e. the PPT
or Peres-Horodecki criterion is in general a necessary but not a sufficient condition.\cite{Peres96} Therefore a state is certainly entangled, whenever one of the eigenvalues of $\rho^{T_1}$ is negative.

By solving a cubic and three quadratic equations the eigenvalues of $\rho^{T_1}$
are obtained in terms of magnetizations, two-point, triad and quad correlation functions.
The expressions are too lengthy and we have not brought them here.
However their forms are simplified by considering other symmetries.

{\it States with translation, reflection and spin-flip symmetries} - At the simultaneous presence of the translation, reflection and spin-flip symmetries only the following eigenvalues could be negative.
\begin{eqnarray}
\no&x_1=K_+\pm G_-,~~ x_2=K_-\pm G_+,~~ x_3=-(r^{1/3}+\frac{r'}{3}),\\
&x_4=\frac{1}{8}\left(M_{-}+P_{-}-\sqrt{(M_{-}-P_{-})^2-16(\Im  q_{zz^+}^{--})^{2}}\right).
\end{eqnarray}
where $r'(<0)$, $r$, $N_{\pm}$, $K_{\pm}$, $G_{\pm}$ and $P_{\pm}$ are functions of quad correlations (see appendix) and $\Im$ denotes the imaginary part.
The anisotropic XXZ spin-$1$ models in a transverse magnetic field have these symmetries.
By varying the non-commuting magnetic field the system experiences different phases in which the correlation functions and thus the negativity  behaves differently in each phase.

{\it States with parity and $U(1)$ symmetries} - For states with parity and $U(1)$ symmetries, among $6$ eigenvalues of the three quadratic equations only the following eigenvalues
\begin{equation}
x_1=\frac 14\left[A_--\sqrt{A_+^2+4|q_{z-}^{z+}|^2}\right],~~
x_2=\frac 14\left[B_--\sqrt{B_+^2+4|q_{z-}^{z+}|^2}\right]
\end{equation}
could be negative, where $A_{\pm}=[q_{zz^+}^{z^-z^+}\pm q_{z^-z^+}^{zz^-}]$, $B_{\pm}=[q_{z^-z^+}^{zz^+}\pm q_{zz^-}^{z^-z^+}]$. The other eigenvalues are given by the standard solution of the cubic equation with coefficients $a$, $b$, and $c$ (see appendix) where $a$ is bounded as $-1\leq a\leq 0$ and the conditions to have a negative eigenvalue have been discussed in Fig.(\ref{Neg-val}).

{\it States with parity, spin-flip and $U(1)$ symmetries} - Finally, for PI states with additional spin-flip, $U(1)$ and translational symmetries, the following eigenvalues could be negative:
\begin{eqnarray}
x_1=\frac 18\left(D_+\pm\sqrt{D_-^2+32[q_{z-}^{+z}]^2}\right),~~
x_2=\frac{P_-}{4},~~x_3=K_{\pm}.\label{ptsf}
\end{eqnarray}
where $D_{\pm}=P_+\pm 4q_{z^+z^-}^{z^+z^-}$.
These results give the negativity of the anisotropic XXZ spin-$1$ models in a longitudinal magnetic field. Adding a single site uniaxial anisotropy to this model does not break the symmetry and the negativity is given by the negative eigenvalues of (\ref{ptsf}). The anisotropic XXZ spin-$1$ in a longitudinal magnetic field with single site anisotropy is used to describe the spin super-solid and spin-flop phases.\cite{Seng07,Ross11}
Meanwhile $NiCl_2-4SC(NH_2)_2(DTN)$\cite{Padu04} is a quantum spin-1 chain with strong easy-plane anisotropy and a new candidate for the BEC of the spin degrees of freedom. Finding the negative eigenvalues of (\ref{ptsf}) the entanglement of the BEC state is determined.

\section{Summary}
 In this paper we have studied the bipartite entanglement, as quantified by negativity, between two spins in two general classes of spin system, the parity-invariant (PI) heterogeneous
spin-($1,\frac 12$) system and parity invariant homogeneous spin-$1$ models. 
All the information needed to analyze bipartite entanglement is contained in the two-qubit reduced
density matrix, obtained from the wave function of the state after all the spins except those at positions $i, j$
have been traced out. The resulting reduced density matrix represents a mixed state of a bipartite system for which a good deal of work has been devoted to quantify its entanglement.
To see the effects of collective behaviors of particles on the bipartite entanglement we have expressed the reduced density matrix and the negativity in terms of spin correlation functions.
We have shown explicitly that the triad and quad correlations like $\la S_zS_{\pm}s_z\ra$ and $\la S_zS_-s_zs_{\pm}\ra$, which demonstrate the quadrupole-single spin and the quadrupole-quadrupole correlations, are emerged in the reduced density matrix and the negativity.
We have also found that in mixed spin-$(1,\frac 12)$ and spin-$1$ systems,
in addition to magnetizations and two-point correlations,
the information coming from the quadrupolar operators ($S_zS_\pm$ and $S_\pm S_z$) play an essential role in the pairwise entanglement. These correlations do not appear in the spin-$\frac 12$ models.
Utilizing our results it has been also achieved the entanglement of several models of interest with higher reflectional, translational,
spin-flip and $U(1)$ symmetries.

The effects of the triad and quad correlations on the other useful quantities such as entanglement entropy, quantum fidelity will be investigated in future.

\section{Acknowledgments} The authors are grateful to Andreas Osterloh for fruitful comments and suggestions.
We also acknowledge Saeed Abedinpour and Reza Sepehrinia for reading carefully the manuscript.

\section{Appendix}
\begin{eqnarray}
\no&N_{\pm}=q_{zz}^{zz}\pm g_{zz},~~ K_{\pm}=\frac 12[q_{zz}^{z^-z^+}\pm q_{z-}^{z+}],~~G_{\pm}=\frac 12[q_{z-}^{z-}\pm q_{--}^{z^+z^-}],\\
\no&P_{-}=N_+- q_{--}^{++},~~ M_{-}=N_-- q_{--}^{--},~~r'=-\frac{1}{4}(L_++4q_{z^+z^-}^{z^+z^-}),\\
\no&r=\frac{1}{96}\bigg[L_+\bigg([\Re q_{zz^+}^{--}]^2-\frac 14(P_++4q_{z^+z^-}^{z^+z^-})M_+-q_{z^+z^-}^{z^+z^-}
P_++2([q_{z-}^{+z}]^2+[q_{z-}^{-z}]^2)\\
\no&-4[q_{z^+z^-}^{z^+z^-}]^2\bigg)+8q_{z^+z^-}^{z^+z^-}
\bigg([q_{z-}^{+z}]^2+[q_{z-}^{-z}]^2+\frac 14P_+M_+-[\Re q_{zz^+}^{--}]^2\bigg)\\
\no&-6\big([q_{z-}^{+z}]^2M_++[q_{z-}^{-z}]^2P_+\big)
-24q_{z-}^{+z}q_{z-}^{-z}(\Re q_{zz^+}^{--})
+\frac{1}{18}(L_++4q_{z^+z^-}^{z^+z^-})^3\bigg].
\end{eqnarray}

Cubic equation:  
\begin{eqnarray}
\no&ax^3+bx^2+cx=0,\\
\no&a=-\frac{N_+}{2}-q_{z^+z^-}^{z^+z^-},\\
\no&b=\frac{1}{16}(q_{zz^+}^{zz^+}q_{zz^-}^{zz^-}-|q_{--}^{++}|^2
+8[q_{z^+z^-}^{z^+z^-}N_+-|q_{z-}^{+z}|^2]),\\ \no&c=\frac{1}{16}(q_{z^+z^-}^{z^+z^-}[|q_{--}^{++}|^2-q_{zz^+}^{zz^+}
q_{zz^-}^{zz^-}]-2\Re(q_{--}^{++}[q_{z+}^{-z}]^2)+2|q_{z-}^{+z}|^2
N_+).
\end{eqnarray}


\begin{thebibliography}{0}
\bibitem{quad} The operators $S_zS_{\pm}$ and $S_{\pm}S_z$ are written in terms of quadrupoloar operators as $S_zS_{\pm}=\frac 12(Q_{ib}\pm iQ_{ic}\pm S_{\pm}), S_{\pm}S_z=\frac 12(Q_{ib}\pm iQ_{ic}\mp S_{\pm})$
\bibitem{spin-flip} The spin-flip symmetry is defined as ${\cal F}S(s)_z{\cal F}=-S(s)_z$, ${\cal F}S(s)_{\pm}{\cal F}=S(s)_{\mp}$ where ${\cal F}$ is the spin-flip operator.
\end{thebibliography}
\end{document}